**Design and construction of a novel tribometer with on-line topography and wear measurement**

Spyridon Korres[1,2] and Martin Dienwiebel[1,2]

[1]Karlsruhe Institute of Technology, IZBS, Kaiserstr. 12, 76131 Karlsruhe, Germany

[2]Fraunhofer-Institute for Mechanics of Materials IWM, Wöhlerstr. 11, 79108 Freiburg, Germany

**Abstract**

We present a novel experimental platform that links topographical and material changes with the friction and wear behavior of oil-lubricated metal surfaces. This concept combines state-of-the-art methods for the analysis of the surface topography on the micro- and nano-scale with the online measurement of wear. At the same time, it allows for frictional and lateral force detection. Information on the topography of one of the two surfaces is gathered in-situ with a 3D holography microscope at a maximum frequency of 15 fps and higher resolution images are provided at defined time intervals by an atomic force microscope (AFM). The wear measurement is conducted on-line by means of radio nuclide technique (RNT). The quantitative measurement of the lateral and frictional forces is conducted with a custom-built 3D force sensor. The surfaces can be lubricated with an optically transparent oil or water.

The stability and precision of the setup have been tested in a model experiment. The results show that the exact same position can be relocated and examined after each load cycle. Wear and topography measurements were performed with a radioactive labeled iron pin sliding against an iron plate.

**1 Introduction**

For the better understanding of fundamental processes in tribology it is important to characterize the interface, which is formed by the two sliding bodies and a possible lubricant. Because this is experimentally very challenging, tribologists often have to rely on simulations. Remarkable advances in this field have been made in the last decades [1, 2]. From simple contact cases to lubricated surfaces, simulations cover a wide range of conditions. Despite these advances, scientists are still in need of scientific experimental results in order to continue delving deeper into the phenomena observed.

A pin-on-disk tribometer and a topographical microscope can be considered common assets of a tribological experiment. Nevertheless, in the vast majority of such experimental procedures, the samples are subjected to several cycles of frictional loads prior to being examined with a microscope. Therefore, they provide no information about the dynamic processes that occur during these cycles. Normal and lateral forces can be recorded during the experiment, but topography and wear are only observed afterwards.

An interesting approach towards studying sliding surfaces *in-situ* is observing one of the two surfaces through the body it is sliding against. This requires that at least one of the two bodies is transparent [3, 4]. The main advantage of this method is that the images acquired, are images of the actual contact area. On the other hand, transparent materials are rarely encountered in demanding tribological systems, so this method poses great limitations to the variety of samples that can be used.

Recently, an effort was made to monitor the wear scar topography after each load cycle. The instrument utilized for these experiments comprises a linear positioner, a white-light interferometer (WLI), and a pin [5]. This instrument can even conduct experiments in different gas atmospheres in a closed chamber; however, it is limited to a linear reciprocating mode and cannot monitor surface changes under lubrication.

While the measurement of *in-situ* topography in a tribological experiment is still being developed, wear particles can be measured using commercial systems. The highest resolution is currently achieved by the so-called "radionuclide-technique" (RNT) [6]. This method involves labeling of one of the two surfaces

(or the entire body) with a radio-tracer and the measurement of wear particles carried away by a lubricant in circulation with a gamma spectrometer. The precision it offers is unique for the simultaneous monitoring of applied forces and wear.

## 2 Experimental

### 2.1 Principle of function and instrumentation

Here we introduce a new concept for the on-line monitoring of the topography and wear measurement of lubricated systems. This is achieved by means of a novel tribometer that consists of state-of-the-art instruments for planar positioning, surface topography, wear measurement and force sensing.

A high precision planar positioning system provides translational freedom on the xy plane. A PPS 200-4 system of Tetra GmbH, Germany, was employed for this purpose. The device has a range of 200 mm×200 mm and reaches a maximum speed of 500 mm/s with ±1 μm precision. The position is monitored at a 20 nm resolution.

A tank which holds a flat plate-sample is placed on top of the positioner which drives it along the desired path. The tank is connected to an RNT (Zyklotron RTM 2000 of Zyklotron AG, Germany) apparatus performing on-line wear measurements. Three measuring devices are attached in three fixed positions over the sample: a force sensor, a holographic microscope and an atomic force microscope (AFM).

In order to measure forces, a customized setup was built, which includes three SKL1417-IR (Tetra GmbH, Germany) fiber optic sensors (FOS). The holographic microscope used is a DHM$^{®;}$ R1000 series (Lyncée Tec SA, Switzerland) in a customized case that fits the present instrumental layout. The objective lenses employed are a 50× oil immersion and a 10× dry. This device can acquire images at a maximum frequency of 15 fps. The software of this instrument was adjusted for the optical properties of the lubricant, in this case transparent poly alpha olefin (Fuchs Petrolub AG, Germany). The AFM utilized here is an ULTRA Objective (Surface Imaging Systems / now Bruker AXS Microanalysis GmbH, Germany) with a scanning range of 80×80 μm$^2$ and an additional kit for immersion measurements.

As the sample moves, the sensors are focused at different positions on the sample surface. The relative position of all instruments is designed in such a way, that all devices have access to a specific part of the path. The schematics of this concept are shown in Fig. 1. A pin is attached at the bottom of the force sensor. With a closed path without angular geometries, the pin can perform smooth cycles without losing contact with the plate-sample. Part of the wear track left behind can be examined with the other two sensors by pausing the motion very shortly. This allows for acquisition of surface topography images after each cycle at the exact same position every time.

### 2.2 Mechanical setup

The base of the tribometer is a granite plate that stabilizes the structure and minimizes height variations. The positioning system is placed directly on the plate and holds the oil-tank. The plate sample is located inside the oil-tank with a polished surface facing up. Two bridges, one made of granite and another one of aluminum profiles, hold the rest of the sensors above the sample. The granite bridge is necessary in order to reduce vibrations that affect the AFM and force sensor measurements. On the other hand, the holographic microscope is less sensitive to vibrations, so a plain aluminum construction is sufficient. The layout of the AFM, force sensor and holographic microscope over the sample matches the one described in the instrumental concept (Fig. 1).

### 2.3 Oil circuit

The oil circuit connects the tribometer with the on-line wear measurement system. A diagram of the oil circulation is shown in Fig. 2. The oil-pump and the radioactivity sensor are located in the RNT apparatus. A bypass circuit is controlled by two chokes, limiting the oil flow at the inlet and outlet of the oil container / reservoir. Depending on the volume of the liquid in the reservoir, the total oil volume varies between 1 and 2 L.

The sample is placed in an oil container; however, it remains above the oil level, as shown in Fig. 3. The lubricant flows onto the sample and carries the particles off the surface. Continuous flow replenishes the reservoir. The bottom of the container is inclined in order to reduce the volume of oil in the reservoir. This improves the resolution of the wear measurement and reduces the inertia of the positioning system. The total mass of the container and the sample without oil is between 7 and 10 kg. The positioning system sets a limit of 150 N normal force. Finally, a tube at the lowest point of the container leads the oil back to circulation.

**2.4 Force sensor**

The force sensor block consists of an up-scaled version of the so-called "tribolever" [7] design, three fiber optic sensors and a steel scaffold holding them together. Because of its small size, the original tribolever comprises a pyramid on top to support the displacement measurement mirrors. In our design, the pyramid is replaced by a cube. Three adjacent faces of this cube are occupied by mirrors: two sideways (x, y) and one on top (z). The fiber optic sensors are facing against the mirrors and measure displacements of the cube in the x, y and z direction. Finally, a pin is screwed at the bottom of the cube. Sketches of the force sensor block can be found in Fig. 4.

As mentioned above, the up-scaled version of the so-called "tribolever" was designed for the measurement of the normal and both lateral forces. The main reason for selecting this geometry is that it provides equal sensitivity for both lateral directions while remaining relatively insensitive to torsional forces. The selected material was steel, with a Young's modulus of 210 MPa. We have examined several variants with different aspect ratios of the four legs using finite element analysis.

Two important factors that had to be considered for the up-scaling were the applied forces and the resulting displacements. Our aim was to optimize the sensor for conventional tribological systems, in which case the frictional force is only a fraction of the normal force. Therefore, the ratio of the spring constant in the z direction ($k_z$) over either one of the lateral directions ($k_x$, $k_y$) should be of the same order of magnitude as the friction coefficient. The applied normal force should result in a pressure that allows for running in. The desired contact area for the present instrument is in the order of tens of mm$^2$. Finally, the displacements should be suitable for the detectors (the fiber optic sensors employed for this apparatus have a range of about 800 μm).

Considering all the above, we concluded in the geometry shown in Fig. 5.

To facilitate a wider range of force measurements, two separate sensors were constructed. These have different width and thickness of the four legs (A and B as shown in Fig. 5) which result in different spring constants. The dimensions and constants of the two sensors are summarized in Table 1.

In order to achieve high geometrical precision, the force sensor was machined by means of spark erosion cutting. A 10 mm wide border of the initial plate was preserved around the legs to support and stabilize the structure as shown in Fig. 4.

**2.5 Samples geometry**

Two samples are required for a tribological experiment with this equipment. A tablet shaped sample that is fixed at the tip of the pin and a plate sample that is placed on the positioning stage.

The pin consists of a screw adjusted to the bottom of the force sensor and a spherical sample holder (Fig. 6). The sample holder allows for rotation with a maximum angle of 9 °. In contact, this enables the tablet shaped sample to be aligned parallel to the tablet surface. the diameter of the sample is 2 mm and the height 1.5 mm. The edge around the circular contact area is tapered.

**2.6 Software and control of the instrument**

In order to perform on-line measurements it is necessary to synchronize all devices. Fig. 7 shows the schematic of the software control. Positioning and force monitoring are performed via a LabView (National Instruments Corporation) program. The connection to the positioning stage is established via TCP. A NI PXI-6120 S Series card (National Instruments Corporation) records the three analog signals (x, y and z) of the force sensor. Separate computers control the optical and holographic microscope as well as the RNT. A TCP connection establishes the communication to the computer which is directly attached to the holographic microscope. This enables the remote control of the microscope and consequently the automation of the acquisition.

The geometry of the sliding path of the planar drive is programmed in a G-code similar manner. The current position is read every 80 µs. The z-position is adjusted according to the current value in order to maintain the normal force constant. Once the objective lens of the holographic microscope has reached a predefined position, the stage pauses and the acquisition of a 3D image is triggered.

3D images gathered during the experiment are organized in frames, which provide an animated view of the surface and graphical representation of forces and friction coefficients. The values shown at these graphs are calculated via interpolation for the position examined with the microscope. The two closest measure samples to this position are used for the interpolation. An example is shown in Fig. 8.

## 3 Calibration and Testing

### 3.1 AFM

As the positioning system readjusts its position dynamically during operation, fluctuations of a constant position may occur. These may affect the stability of the AFM measurements. To ensure the stable function of the atomic force microscope, several tests were performed with dry and water immersed samples. An example of a copper surface is shown in Fig. 9. The surface was examined with the positioning system turned off, as well as with the positioning system turned on and the AFM working in both dry and immersion mode. Given that no additional shielding or damping was applied to improve the scanning process, the resulting images have acceptable levels of noise and the surface structure is well recognizable.

### 3.2 Position calibration

In order to examine a single position on a flat sample with all three sensors (force sensor, holographic microscope and AFM), their relevant position has to be determined. This task was performed in two steps. First, a sphere was attached to the force sensor, in the same position where the tablet shaped sample is normaly located. With this setup, a cross was engraved on the plate sample at a given position of the stage $(x_0, y_0)$. The center of this cross was then located with the holographic microscope by moving the stage to a new position $(x_1, y_1)$. A patterned silicon sample with numbered blocks was then placed on the same position on the plate sample and an image was acquired with the holographic microscope. The stage was moved again until the same position of the patterned sample was relocated with the AFM $(x_2, y_2)$. Images of the patterned sample are shown in Fig. 10, which demonstrates the precision with which the instruments can be synchronized.

When designing a path for a tribological experiment it is necessary to program a path that is accessible to the force sensor and the holographic microscope. The simplest case is a line that connects the relative positions of these elements. The slope and offset of this line is calculated as shown below:

$$A = \frac{y_1 - y_0}{x_1 - x_0} \quad (1)$$

$$B = y_0 - A x_0 \quad (2)$$

In such an experiment, if the pin is in contact with the tablet at position $(x_a, y_a)$, which is to be examined, then the acquisition with the holographic microscope will be made at $(x_b, y_b)$:

$$y_b = y_a + y_1 - y_0 \quad (3)$$

$$x_b = x_a + x_1 - x_0 \quad (4)$$

And for any other position (x,y) a random y position is coupled to:

$$x = \frac{y - (y_a - Ax_a)}{A} \quad (5)$$

The path is further adjusted accordingly to include the position of the stage for an AFM measurement.

### 3.3 Example of an experiment

An extended test was performed to try the instrument in experimental conditions. For this purpose, two samples of pure iron (>99.9 % purchased by Goodfellow GmbH, Germany) were prepared: a tablet and a plate shaped sample. The edges of the tablet were tapered to avoid indentation of the sample's side in the counter-face and the contact surface was polished. The contact area was round in shape with a diameter of approximately 1.9 mm (area of 3.61 mm$^2$). The sample was labeled using neutrons in a nuclear reactor. The specific activity of the sample was 17,21 kBq (Fe-59).

A linear path connecting the force sensor and the holographic microscope was programmed. The stage moved in a reciprocating motion at 15 mm/s. The normal force was set to 10 N (nominal pressure of 2.7 MPa) and the length of the line was 120.03 mm. A total of 5450 cycles were recorded, which corresponds to a total sliding length of 654.17 m. The experiment was performed using poly alpha olefin as lubricant.

During the first 100 cycles, the precision in the positioning and force regulation were monitored. Since the programmed path was a straight line, the deviation of the stage position from the theoretical line was calculated by their perpendicular distance. The results are shown in Fig. 11. The average deviation calculated was 18 nm. This result proves that the path is followed precisely enough to ensure that the pin remains on the same track throughout the test. This deviation does not affect the acquisition of the 3D images, as it is lower than the resolution that the holographic microscope provides. On the other hand, the dynamic stabilization of the normal force showed greater deviations. As the plate sample was not polished, the roughness of the surface had an impact on the normal force regulation, which gave rise to a standard deviation of 31 % of the set force.

Fig. 12 gives an overview of the entire experiment. In the beginning, both the friction coefficient and the wear rate increase. In this phase of the experiment, no significant change is observed in the images acquired by the holographic microscope. After 600 min, the left part of the image starts changing. This happens gradually, and expands to further asperities coming into contact in the rest of the window observed with the microscope. In this phase, the friction coefficient and the wear rate start to stabilize. Deeper scratches in directions other than the sliding direction slowly fade away. After 1200 min these scratches are no longer recognizable and the wear scar has expanded on the entire surface. A total value of 786.5 nm wear was calculated as total wear at the end of the experiment.

It is interesting to note, that the wear rate and topography changes seem to be correlated, while the friction force evolves independently.

### 4 Summary

The presented tribometer combines modern techniques that enable the monitoring of the surface of one of the two tribological counter-faces, while measuring normal, frictional forces and wear. It's major breakthrough is the fact that it's function is not limited to dry or transparent systems. On-line topography in lubricated tribosystems can help understand the phenomena that occur in several applications in greater depth.

Thorough tests have been performed to determine the precision and stability of function. These show, that one single position of the wear scar can be relocated and examined after each cycle of an experiment.

This allows for dynamic observations of the changes that occur. The resolution offered by the holographic microscope is sufficient to provide information on the plastic deformation that occurs at specific asperity junctions. Further detail can be obtained by scanning the same position with the AFM. At the same time, the behavior of the friction coefficient and wear rate can be displayed throughout the experiment. As these usually exhibit a non-linear behavior against time, interesting changes can be coupled with observations on the surface of the sample.

In this paper we have presented the results of a simplified experiment with a linear path and iron samples. Nevertheless, this tribometer can be utilized for experiments that demand greater geometric complexity. In addition, as there are few requirements for the materials of the selected samples, several material combinations can be tested. One important limitation is connected to the on-line wear measurement: the samples must contain isotopes that can be activated.

This work was funded by the German Science foundation DFG (under the project number DI 1494/1-1).

The authors also wish to gratefully thank Matthias Scherge, as well as Olaf Mollenhauer and Matthias Neuland of Tetra GmbH and Yves Emery of Lyncée Tec SA for their assistance and discussions.

**Tab. 1:** Dimensions (A: thickness and B: width as shown in Fig. 5) and spring constants ($F_n$: normal force and $F_f$: frictional force) for the two force sensors $S_1$ and $S_2$.

| Sensor | $k_f$ [N/mm] | $k_n$ [N/mm] | A [mm] | B [mm] |
|---|---|---|---|---|
| $S_1$ | 9.45 | 20.90 | 0.5 | 1.0 |
| $S_2$ | 288.25 | 244.40 | 1.0 | 3.0 |

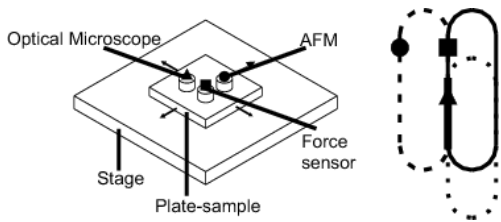

**Fig. 1:** According to the concept of the present instrument, the sample has translational freedom on the xy plane, limited by the size of the stage. The force sensor with the pin (square), the optical / holographic microscope (triangle) and the AFM (circle) are attached at fixed positions over the sample. For a race-track shaped motion: the dashed, dotted and thin continuous lines on the right show the paths on the sample accessible to the AFM, optical microscope and force sensor respectively. The thick continuous line is accessible to all sensors.

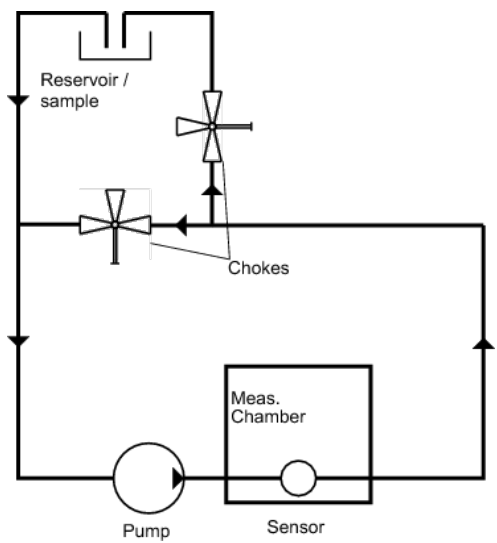

**Fig. 2:** Oil circulation diagram. The pump, is followed by a radioactivity sensor. Both are located in the RNT instrument. Outside this apparatus, the flow is split to access the oil container / reservoir and forms a bypass circuit. Two chokes control the flow in each direction. The sample is located in the reservoir area.

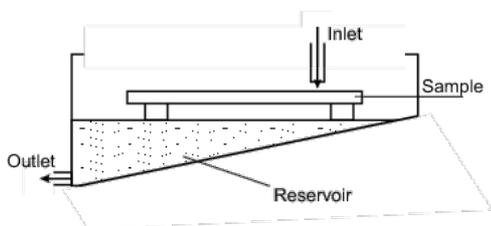

**Fig. 3:** The sample is located above the oil reservoir. The lubricant flows onto the sample and is collected from the lowest point of the reservoir.

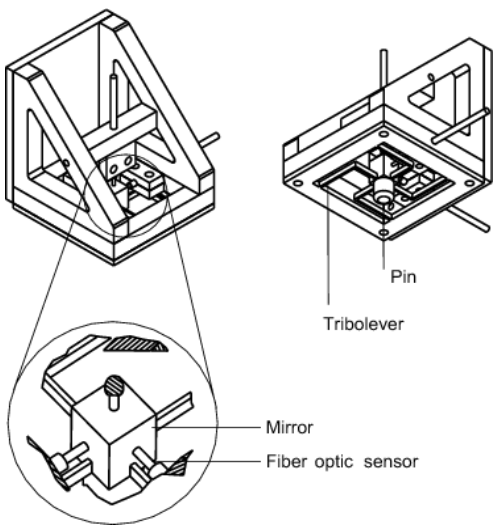

**Fig. 4:** A customized scaffold holds the force sensor. The pin is screwed at the bottom face of the central cube, while three mirrors are attached on the top and two adjacent side faces. Fiber optic sensors are fixed against them and measure the x, y and z displacement.

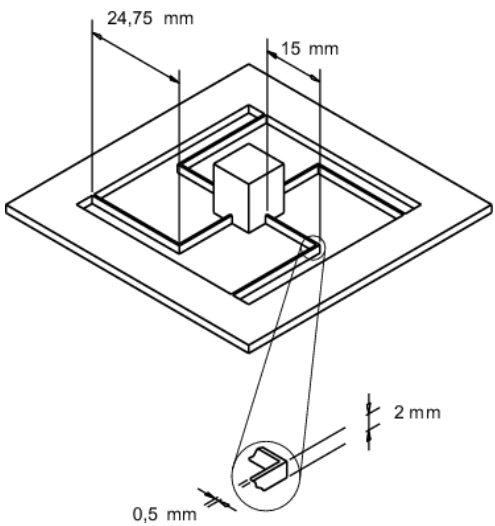

**Fig. 5:** The force sensor is an up-scaled version of the "tribolever". The dimension and aspect ratio of the legs have been optimized for macroscopic tribological experiments.

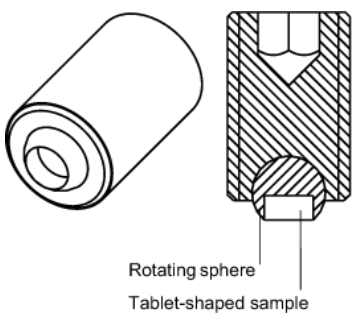

**Fig. 6:** The screw-pin with a rotating sphere. The tablet shaped samples are inserted in the sphere, which allows a max. of 9 ° rotation.

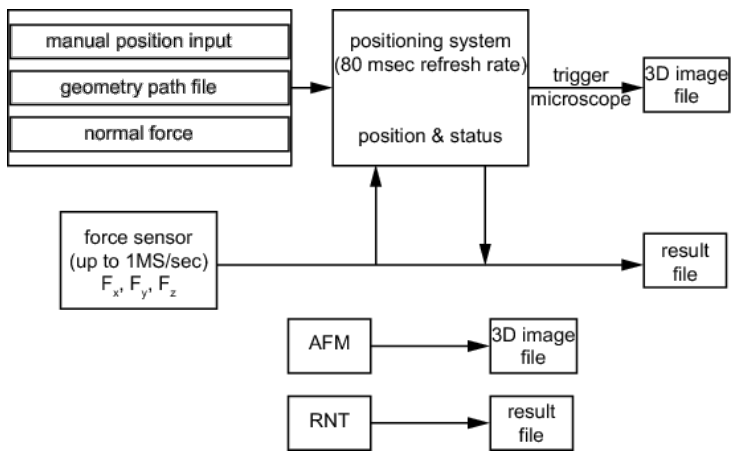

**Fig. 7:** Software and control diagram.

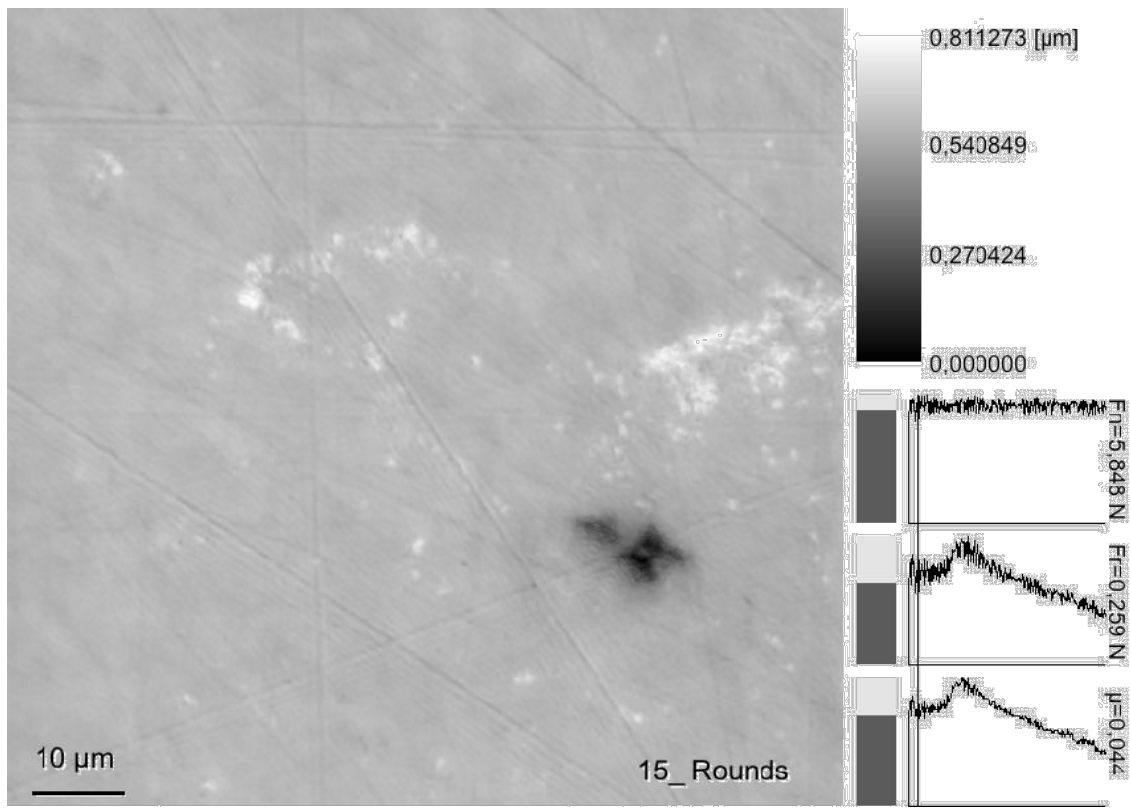

**Fig. 8:** 3D representation of a Cu sample surface (left) with a grey scale for the z axis (top right) as well as measured forces and friction coefficient for the current cycle (lower right).

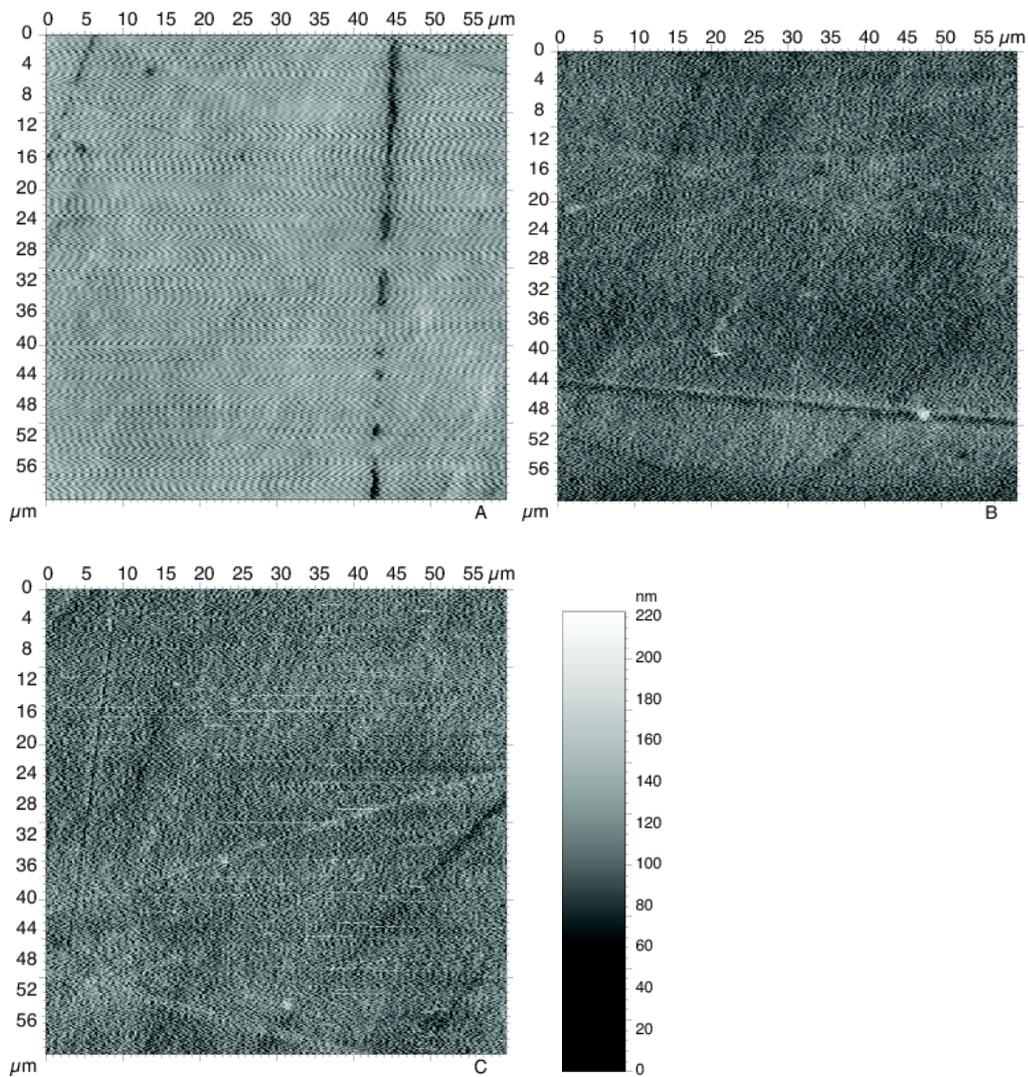

**Fig. 9:** AFM measurements of a polished copper surface in dry mode with the positioning system off (A), in dry mode with the positioning system on (B), and in water immersion with the positioning system on (C).

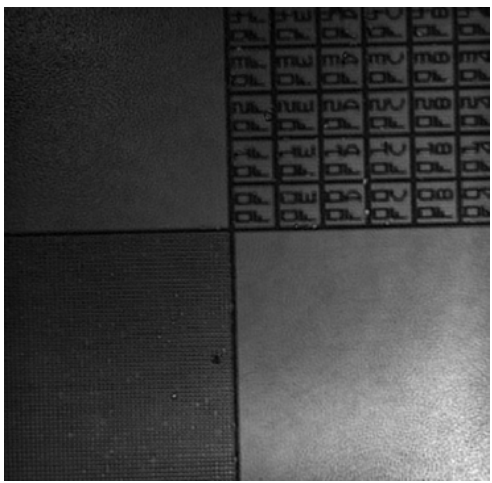

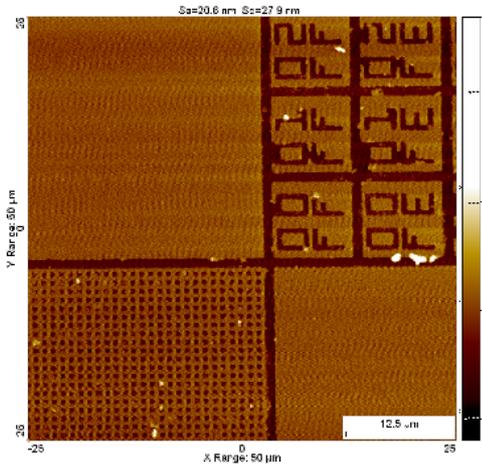

**Fig. 10:** Reference patterned sample with numbered blocks located with the holographic microscope (top) and the same sample relocated with the AFM (bottom) by moving the stage.

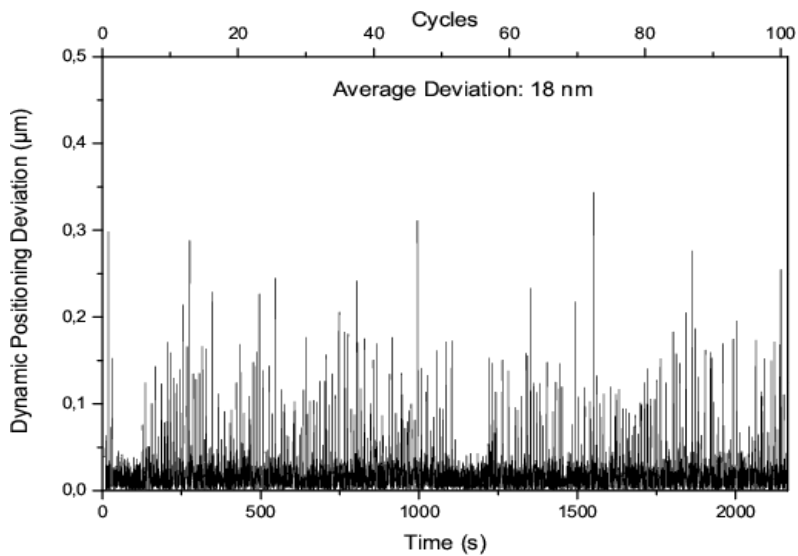

**Fig. 11:** Dynamic perpendicular positioning deviation of the stage during the first 100 cycles of a reciprocal motion with a linear path of 120 mm. The experiment had a set normal force of 10 N with a sliding speed of 15 mm/s of an iron pin against an iron plate with PAO lubrication.

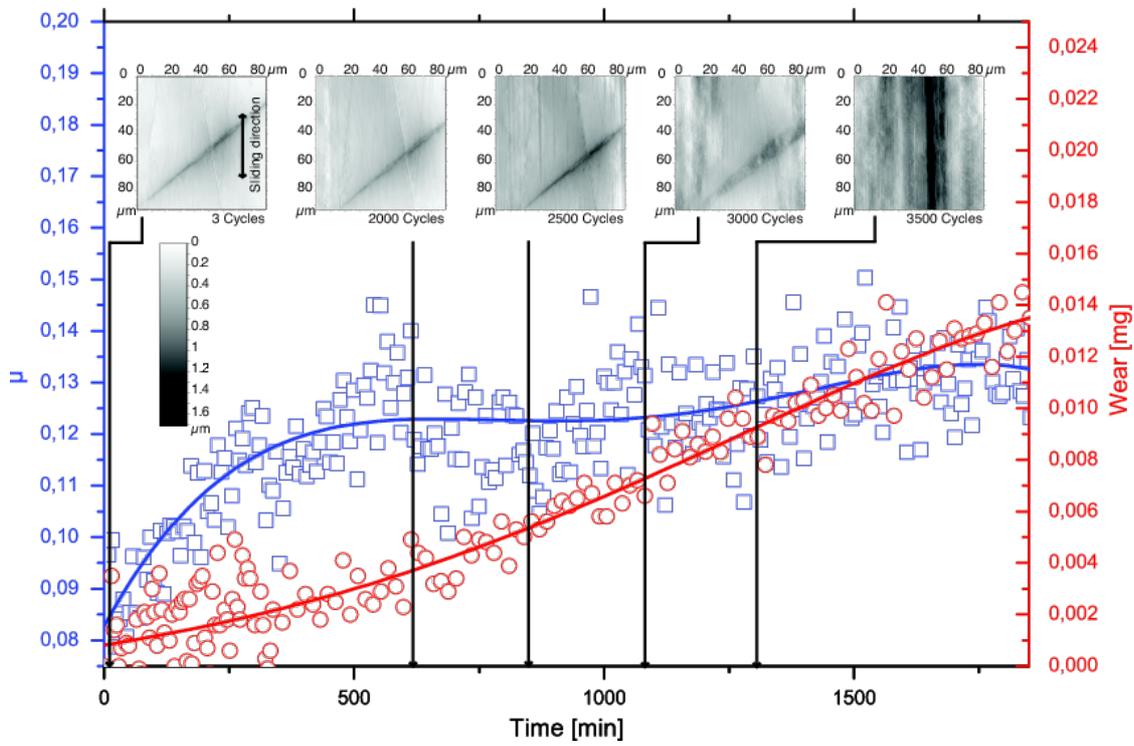

**Fig. 12:** Evolution of the topography, friction coefficient (open squares) and measured wear (open circles) during frictional load (normal force) of 10 N and sliding speed of 15 mm/s of an iron pin against an iron plate with PAO lubrication.